\documentclass[twocolumn, trackchanges]{aastex61}

\usepackage{rotating}
\usepackage{ulem}
\usepackage{color}

\newcommand{\msunh}{\>h^{-1}\rm M_\odot}

\newcommand{\mpch}{\>h^{-1}{\rm {Mpc}}}
\newcommand{\kpch}{\>h^{-1}{\rm {kpc}}}

\newcommand{\magarcsec}{ \ \rm{mag} \ arcsec^{-2}}

\def\gtsima{$\; \buildrel > \over \sim \;$}
\def\ltsima{$\; \buildrel < \over \sim \;$}
\def\gta{\lower.7ex\hbox{\gtsima}}
\def\lta{\lower.7ex\hbox{\ltsima}}

\received{}
\revised{}
\accepted{}
\submitjournal{ApJ}

\shorttitle{Investigation of intra-cluster light fraction}
\shortauthors{Tang et al.}

\begin{document}

\title{An Investigation of intra-cluster light evolution using cosmological hydro-dynamical simulations}

\correspondingauthor{ Lin Tang,  Weipeng Lin, Xi Kang}
\email{tangl08@shao.ac.cn, linweip5@mail.sysu.edu.cn, kangxi@pmo.ac.cn}

\author{Lin Tang}
\affil{Key Laboratory for Research in Galaxies and Cosmology, Shanghai
Astronomical Observatory; Nandan Road 80, Shanghai 200030, China}
\affil{University of Chinese Academy of Sciences, 19A, Yuquan Road,
Beijing, China}

\author{Weipeng Lin}
\affil{School of Physics and Astronomy, Sun Yat-sen University,
Xingang West Road 135, Guangzhou 510275, China}
\affil{Key Laboratory for Research in Galaxies and Cosmology,
Shanghai Astronomical Observatory; Nandan Road 80,
Shanghai 200030, China}

\author{Weiguang Cui}
\affil{Departamento de F\'{i}sica Te\'{o}rica, M\'{o}dulo 15,
Facultad de Ciencias, Universidad Aut\'{o}noma de Madrid,
28049 Madrid, Spain}

\author{Xi Kang}
\affil{The Partner Group of MPI for Astronomy, Purple
Mountain Observatory, 2 West Beijing Road, Nanjing 210008}

\author{Yang Wang}
\affil{School of Physics and Astronomy, Sun Yat-sen University,
Xingang West Road 135, Guangzhou 510275, China}

\author{E. Contini}
\affil{The Partner Group of MPI for Astronomy, Purple
Mountain Observatory, 2 West Beijing Road, Nanjing 210008}

\author{Yu Yu}
\affil{Key Laboratory for Research in Galaxies and Cosmology,
Shanghai Astronomical Observatory; Nandan Road 80,
Shanghai 200030, China}


\begin{abstract}
The intra-cluster light (ICL) in observations is usually identified through the surface brightness limit method. In this paper, for the first time we produce the mock images of galaxy groups and clusters using a cosmological hydro-dynamical simulation, to investigate the ICL fraction and focus on its dependence on observational parameters, e.g., the surface brightness limit (SBL), the effects of cosmological redshift dimming, point spread function and {\small CCD} pixel size. Detailed analyses suggest that the width of point spread function has a significant effect on the measured ICL fraction, while the relatively small pixel size shows almost no influence. It is found that the measured ICL fraction depends strongly on the SBL. At a fixed SBL and redshift, the measured ICL fraction decreases with increasing halo mass, while with a much faint SBL, it does not depend on halo mass at low redshifts. In our work, the measured ICL fraction shows clear dependence on the cosmological redshift dimming effect. It is found that there are more mass locked in ICL component than light, suggesting that  the use of a constant mass-to-light ratio at high surface brightness levels will lead to an underestimate of ICL mass. Furthermore, it is found that the radial profile of ICL shows a characteristic radius which is almost independent of halo mass. The current measurement of ICL from observations has a large dispersion due to different methods, and we emphasize the importance of using the same definition when observational results are compared with the theoretical predictions.
    
\end{abstract}




\section{Introduction}
\label{sec_intro}
The concept of intra-cluster light (ICL) or luminous intergalactic matter was  firstly introduced by \cite{Zwicky1951} during his studies on Coma cluster. Most of ICL locates in cluster center and surrounds the brightness centre galaxy or brightest cluster galaxy (BCG). It is thought to be the light from stars, which fills the intergalactic space in dense galaxy environments and bounds to the cluster potential but not to any individual galaxy. Observationally, ICL has been found in local universe such as Coma, Virgo cluster \citep[e.g.,][]{Mihos17, Longobardi13, AG10, Gonzalez07, Mihos05}, in clusters at intermediate redshift \citep[e.g.,][]{Giallongo14, Melnick12,Toledo11} and in some clusters at high redshift \citep[e.g.,][]{Adami13, Burke12}. Using large samples of clusters from low to intermediate redshift, such as SDSS \citep[e.g.,][]{Zibetti05, Budzynski14}, 2MASS \citep[e.g.,][]{LM04} and CLASH \citep[e.g.,][]{Burke15}, one can also obtain ICL fraction by the way of stacking galaxies and center region of galaxy clusters. These observations give opportunities to explore the physical properties of ICL, such as age, metallicity, velocity, color and spatial distribution \citep[e.g.,][]{Mihos16, MT14, Presotto14, AG10, Arnaboldi04}. 

The ICL is now widely regarded as an important component of galaxy cluster. Its accurate determination has important implication on the mass of bright cluster galaxies that may alter the measured shape of stellar mass function or luminosity function at the massive end \citep[e.g.,][]{Li&White2009, Bernardi2013, HeYQ2013, DSouza2015}.  Therefore it can be used as an important ingredient  to constrain the theoretical model of galaxy formation \citep[e.g.,][]{Contini2017}. However, the studies of how ICL fraction evolves with halo mass and redshift are far from conclusive because of different methods used in observations and the difficulty of obtaining data for intermediate or high redshift clusters.

As it is difficult to analyze the formation and evolution history of ICL directly from observational results, more feasible approach is to investigate the problem by using simulations. Inspired by the early work of \cite{Merritt1984} who used numerical simulation to firstly show that ICL was formed by stars stripped from merging galaxies in cluster, numerous works using N-body and hydro-dynamical simulations have been devoted to study the formation and properties of ICL \citep[e.g.,][]{Murante04, Murante07, Willman04, Tutukov07, Sommer-Larsen05, Puchwein10, Dolag10, Rudick06a, Rudick09, Rudick11, Barai09, Cui2014a}. These theoretical studies partly confirmed that the major of ICL is formed by dynamical stripping, and its physical properties vary from different dark matter haloes. However, the ICL fraction drawn from numerical simulation is significantly higher than observations and depends on the dynamical models \citep[e.g.,][]{Puchwein10}. Meanwhile, lack of consistent method to measure ICL also hampered the comparison between simulations and observations.

By using analytical description of  how stars are stripped from member galaxies falling  into a cluster, the ICL can also be estimated from the semi-analytical models based on  the cluster formation history \citep[e.g.,][]{DeLucia04, Martel12, Contini14}. These theoretical studies generally found that ICL fraction is closely related to cluster formation history and it increases steadily with time, with the present fraction varying from $10\%$ to $50\%$ in clusters \citep[]{Murante04}. Considering the multi-parameters used and lack of cluster galaxy population, semi-analytical model can hardly obtain pinpoint statistics of ICL properties.

So far, there are still great discrepancies among results with different methods. For example, \citet{Murante04} found that the massive simulated clusters have a larger fraction of stars in the diffuse light than low-mass ones, while no dependence on halo mass in reported by \citet{Puchwein10}. The difference between these two studies is whether to include AGN feedback in the simulations. 

Indeed,  as one can see that the ICL is the remaining component except for cluster member galaxies,  the main problem and difficulty turn out to be the method to define galaxies within a dark matter halo, especially the brightest central galaxy. For example, in some studies galaxies were defined as distinct stellar groups using the friends-of-friends (FoF) algorithm with an arbitrary linking length. Hence the linking length largely decides galaxy size and the remaining ICL. Improvements have been made to include only gravitational bound stellar particles to galaxies \citep[e.g.,][]{Puchwein10, Dolag10}. This dynamical method looks more physical, however, it is not directly applicable to compare with observational results.

Contrary to the relatively straightforward dynamical method mentioned above \citep[see also][]{Cui2014a}, observational estimate of ICL fraction is much more difficult and  there is no consensus on finding a robust way to identify ICL \citep[e.g.,][]{Zibetti05,  Feldmeier04a}. In the literatures, there are three primary definitions of ICL.

\paragraph*{ ICPNs method}:
Intra-cluster planetary nebulas (IC PNs), intra-cluster red giant branches (IC RGBs) and globular clusters (GCs) can be used as tracers of ICL in observations \citep[e.g.,][]{Longobardi13, Ventimiglia11, Peng11, Castro09, Mihos09, Sand08, Williams07}. This method can provide more accurate measurement of ICL. However, it requires deep observations and is often applied to close targets, such as the Virgo and Coma cluster in local universe.

\paragraph*{SB Profile or $0.25R_{vir}$ method}:
These methods distinguish ICL from cluster galaxies by the difference of their intrinsic properties, such as the kinematic property \citep[e.g.,][]{Cui2014a, Rudick11}, the surface brightness (SB) profile \citep[e.g.,][]{Giallongo14, Melnick12, Jee10}, spatial distribution \citep[e.g.,][]{Demaio15, KB07}, or mass distribution \citep[e.g.,][]{Rudick09} in observation and simulation. These methods are normally used for clusters at low and median redshifts in observation.  

\paragraph*{SB limit method}:
This method directly defines the light from stars as diffuse light which is fainter than a characteristic surface brightness limit (SBL) in the observed images \citep[e.g.,][]{Presotto14, Puchwein10}. This method can be applied to objects at any redshift. It is a simple and straightforward method to estimate ICL fraction if deep optical images could be obtained from observational facilities.
\vspace{10pt}

These above definitions of ICL are quite different and they are often applied to targets  at different redshifts with different observational depth. Therefore, it is not surprising that large discrepancies on the ICL fraction have been found and controversial conclusions have then been made. In some cases, ICL is also called as diffuse (unbound) stellar component (DSC). It is noted that the definition of DSC is a more physical one, while the ICL is not necessary only from the unbound stars, depending strongly on the methods used to define ICL.

Combined with the case that observational works and theoretical studies often use  different definitions of ICL, the comparison between data and model predictions is  complicated and not reliable. To fully understand the formation and abundance of ICL, we need consistent comparisons between observations and theoretical models, which can be achieved only through mock observation, i.e., applying observational definitions of ICL to simulated galaxy clusters. \cite{Cui2014a} used simulated galaxy clusters to compare the fraction of ICL between two different definitions, one is following a physical definition of ICL which seperates the ICL from the BCG through fittings of double velocity dispersion distributions of their star particles, and the other is mimicking observational processing which defines ICL by the SB limit method. They found that the two methods produce ICL fraction with a factor of 2-4 depending on the gas physics implemented in the simulation.

In this work we utilize a cosmological hydro-dynamical simulation to identify ICL in the halos of galaxy groups and clusters. Similar to \cite{Cui2014a}, the SB limit method is applied to define ICL and the mock `galaxy' images are produced, where `galaxy' refers to simulated galaxy. To mimic observation more realistically, we make an improvement to additionally consider the {\small CCD} pixelation, the smoothing effect by point spread function ($PSF$) , as well as the cosmological redshift dimming effect. Basically, our main goal is to investigate the various selections and systematic effects used in the SB limit method when measuring ICL, rather than comparison with current observations (which will still be discussed) or other theoretical predictions. This work can serve as understanding the measured ICL fraction and its dependence on observational selection effects, and the predicted trends can also be tested using future observations. 

This paper is organized as follows. In Section~\ref{simulation} we introduce the simulation data and the methods to produce mock `galaxy' images are given in Section~\ref{method}. We present the main results in Section~\ref{results} and the comparison between ICL from different observations is given in Section~\ref{comparison}. Finally the conclusion and discussion are given in Section~\ref{summary}.

\section{simulation}
\label{simulation}

In this work, we utilize a cosmological simulation run with the massive parallel N-body  code {\small GADGET-2} \citep[]{Springel05}. The simulation is evolved from redshift  $z=120$ to the present epoch in a cubic box of $100 \mpch$ with $512^3$ particles for both dark matter and gas particles respectively. We use a flat $\Lambda{\rm CDM}$ ``concordance'' cosmology with $\Omega_m=0.268$, $\Omega_{\Lambda}=0.732$,  $\sigma_8=0.85$, and $h=0.71$. A Plummer softening  length of  $4.5 {\rm kpc}$ is  adopted in the simulation.  In our simulation each  dark matter particle has a mass of $4.62\times10^8 \msunh$, and the initial mass of  gas particles is $9.20\times10^7 \msunh$ which can  be turned  into two  star particles  later  on. The simulation includes  the processes  of radiative heating  and cooling, star formation,  supernova feedback,  outflows by galactic  winds, and metal enrichment, as well as a sub-resolution multiphase model for the interstellar  medium. This model  is  implemented using smoothed particle  hydrodynamics, and enables us to  achieve a wide dynamic range in simulations of structure formation. The star formation  time-scale in the quiescent  model for star  formation is  directly determined  form observations  of local disc galaxies.  Interested readers  are referred to \cite{SH03} for more details about the treatment of gas physics. The simulation has been used in several studies \citep{Jing06, Lin06, Jiang08}.

From the simulation dark matter halos are firstly  obtained   through  the  standard FoF algorithm with a linking length of 0.2 times the mean particle separation. Only halos with a minimum particle number of 60 will be selected for later  analysis. The virial radius $R_{200}$ \footnote{200  is respecting to the  cosmic critical density at a  given redshift, $\rho_c(z)  = 3H^2(z)/(8 \pi G)$.}  and virial mass   $M_{200}$   of   FoF   halo are   calculated   according   to \cite{Lin06}.  In this  paper, only halos with  $M_{200} \geq 10^{12.5}  \msunh$ are included to make sure that they contain  enough star particles.

\section{Methods}
\label{method}
\begin{figure*} \epsscale{0.57}
  \plotone{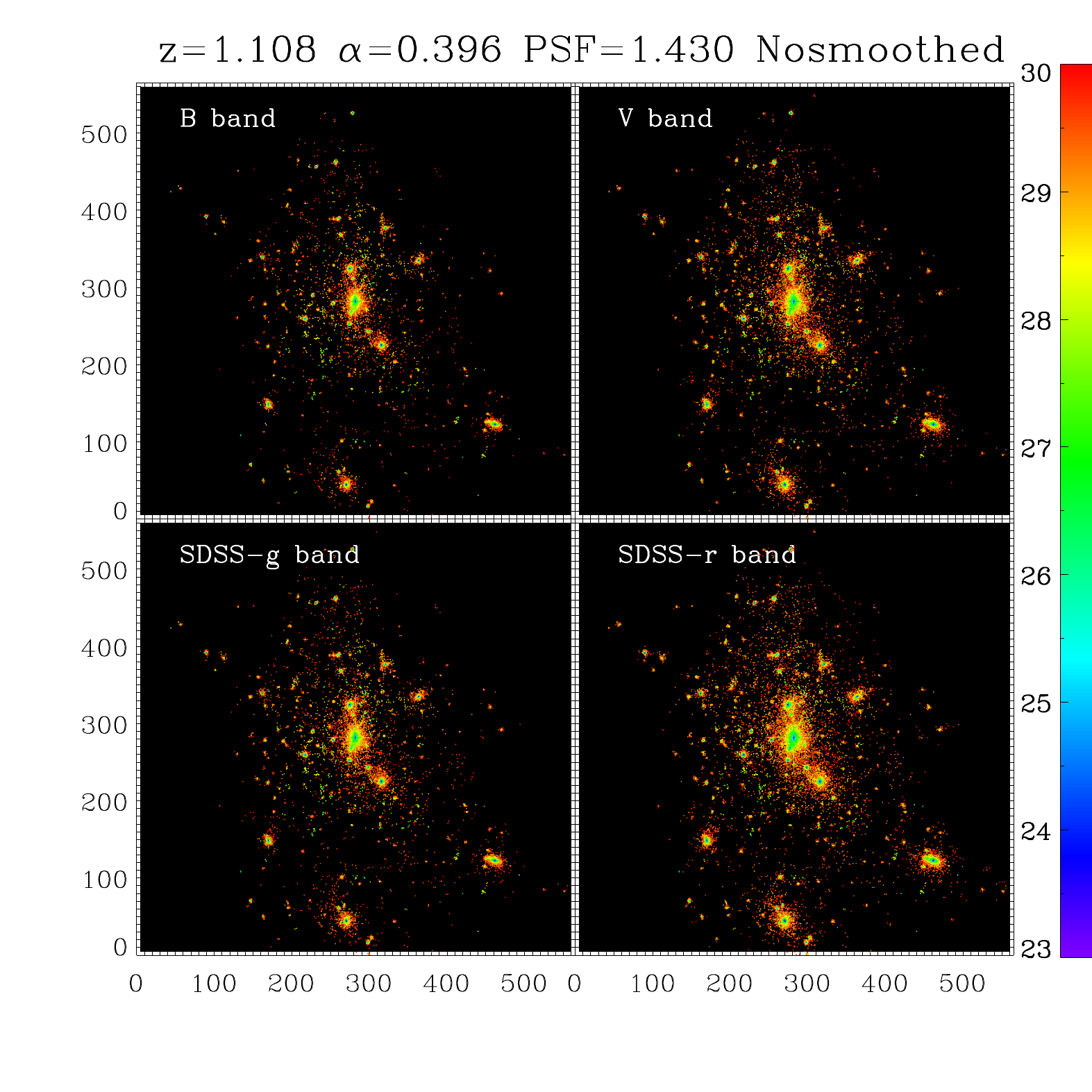}\plotone{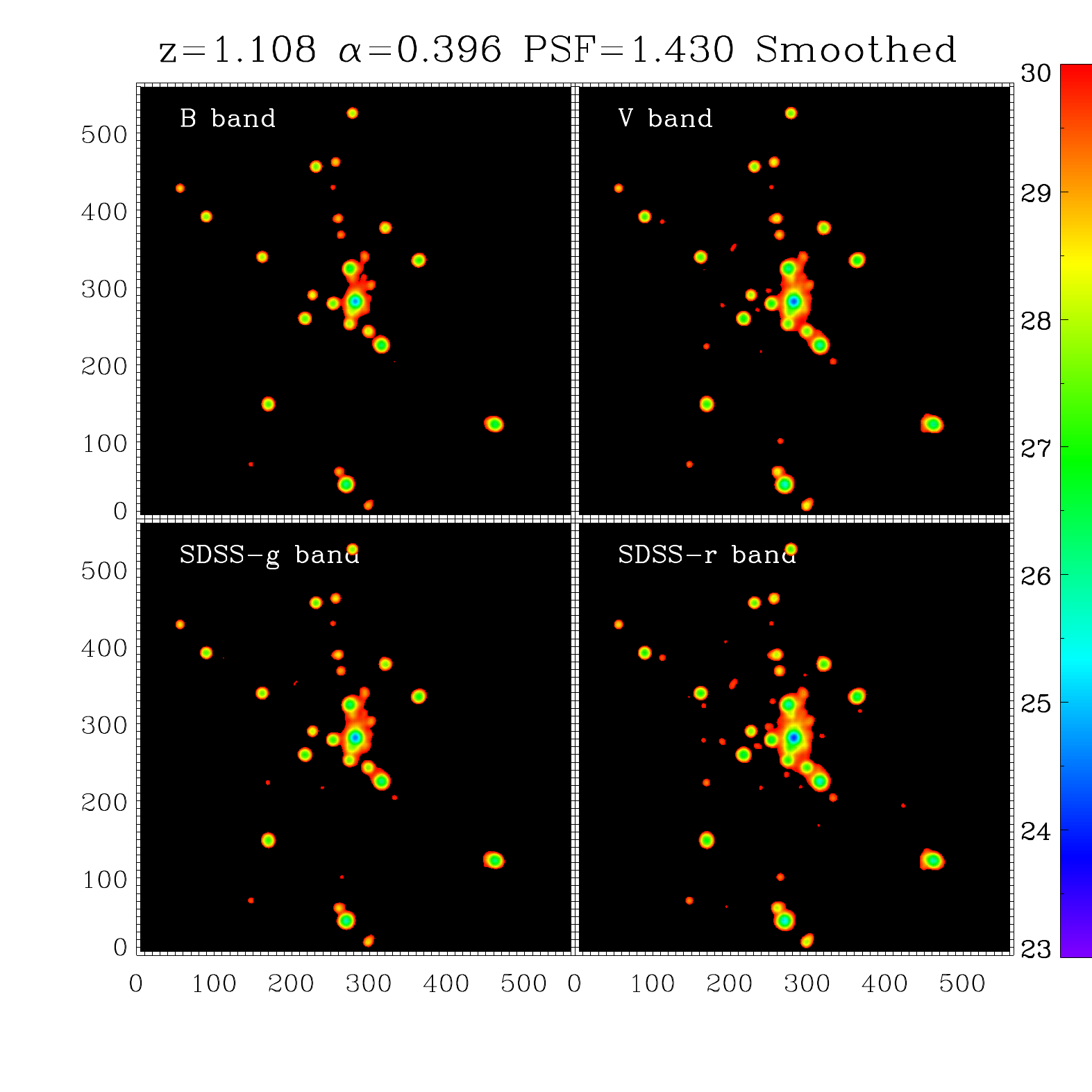}
     \caption{Surface brightness maps for a galaxy cluster at $z = 1.108$ without  $PSF$ smoothing (left panels) and with $PSF$ smoothing (right panels). Each panel shows the surface brightness at four different observational bands, $B$, $V$, SDSS $g$, and SDSS $r$ at rest-frame, from left to right and from top to bottom respectively. The pixel size is $0.396^{''}$ ($D \approx 2.32 \kpch$), $PSF$ width $\omega$ is $1.43^{''}$. Note that we apply the same color gradients for both maps.}
     \label{figure1}
\end{figure*}

To generate the surface brightness maps of simulated galaxies, we follow a similar procedure as in \cite{Cui2011, Cui2014a}. Each star particle of  the FoF group is  treated as a  Simple Stellar Population (SSP)  with age, metallicity and mass given by the corresponding particle's properties in the simulation. The \cite{BC03} model is used to calculate the spectral energy distribution of the model galaxy and a Chabrier IMF (Chabrier 2003) is assumed. Spectral templates from 6 metallicities (0.0001, 0.0004, 0.004, 0.008, 0.02, 0.05) with 69 ages (between $10^5$ yr and $10^{10}$ yr) are used. The simulated star formation time $t$ and metallicity $\log Z$ of the star particle are linearly interpolated in the tables to produce the magnitudes at different filters. In this paper the results are presented in the $B$, $V$, SDSS $g$, SDSS $r$ filters which are quoted at rest-frame in the AB magnitude system.

To  generate the  two  dimensional (2D)  surface  brightness maps,  we firstly apply 2D grids for every group out to its virial radius, where each grid is corresponding to a {\small CCD} pixel. Then, at a given redshift $z$, the grid size $D$ and the angular-diameter distance $d_{A}$ is related as,
\begin{equation}
	\label{eq_D}
	d_A=\frac{ D }{ \alpha  }=\frac{ a_0 r }{ 1 + z },
\end{equation}
and luminosity distance $d_{L}$ is given by,
\begin{equation}
	\label{eq_L}
	d_L= a_0 r ( 1 + z ),
\end{equation}
where $\alpha, \ a_{0}, \ r$ are {\small CCD} pixel scale, scale factor at the present time and proper distance, respectively. If not particularly noted, pixel scale $\alpha$ is set to $0.396^{''}$, a typical value for SDSS CCDs. Using the Friedmann Equation, $a_{0}r$ is given by,
\begin{equation}
	\label{eq_ar}
	a_{0} r=\frac{ c }{ H_{ 0 } }\int_{ 0 }^{ z } { \frac{ dz }
	{[ \Omega_{ \Lambda,0 } + \Omega_{ m,0 }( 1+z )^3 ]^{1 / 2} } }.
\end{equation}
For the adopted cosmological parameters, it is found that the angular-diameter distance $d_{A}$ and grid size $D$ have a peak at around $z\sim1.6$ and then decreases at higher $z$, leading to the well-known observational effect of increasing surface brightness of the objects.

With the  estimated grid size $D$,  the star particles  inside the group virial radius are  binned into a 2D mesh on x-y  plane with weights of both either  luminosities or stellar masses.  The surface brightness  for each pixel is given by,
\begin{equation}
	\label{eq_ux}
	\mu_x=-2.5\log\frac{I_x}{L_{\odot,x} \cdot pc^{-2}}+21.572+M_{\odot,x},
\end{equation}
where $I_{x}$ is,
\begin{equation}
	\label{eq_lx}
	I_x=\frac{L_x}{\pi^2 D^2}(1+z)^{-4}.
\end{equation}
Here $x$ indicates different filter bands at rest-frame \citep[cf.][]{Mo2010}, and $M_{\odot,x}$ is the absolute solar magnitude, which are $5.36, \ 4.80, \ 5.12, \ 4.64 \magarcsec$ \citep{BR07} for $B$, $V$, SDSS $g$, SDSS $r$ bands in AB system, respectively. In the above equation, the surface brightness is dimmed  by a factor of $(1+z)^{-4}$ and it is verified that such a dimming  is real  in an expanding universe\citep[e.g.,][]{SL01} . In Section \ref{influ_uv}, we will discuss the results of ICL fraction without this dimming effect.

To mimic the observed image of `galaxies', we need to model the effect of point spread function ($PSF$). The original image is then convolved with a 2D $PSF$ kernel for a typical image survey instrument, for example, the Sloan instrument, which are represented by a $51\times51$ matrix. For groups with a grid numbers less than $51\times51$, we fill with empty grids to meet this grid number. With a given width $\omega$, the $PSF$ kernel is given by 2D Gaussian distribution,
\begin{equation}
	\label{eq_gaussian}
	f(x,y)=\frac{1}{2\pi\sigma^{2}}e^{\left[-\frac{
	(x-\mu)^2+(y-\mu)^2}{2\sigma^2}\right]},
\end{equation}
where $\sigma=\omega/\alpha$ and $\mu=25$. For the results at different bands, both grid luminosity and mass are smoothed by applying this process.

In Figure~\ref{figure1},  we illustrate the rest-frame surface brightness maps in each band with (right panels)  and without (left panels)  the $PSF$ smoothing. This halo is the most massive one from our simulation at $z = 1.108$. An angular pixel size of $\alpha =  0.396^{''}$ and a $PSF$ width of $\omega = 1.43^{''}$ are adopted as fiducial parameters for a reference. Note that there is no particular reason to choose these parameters and they are just taken from the SDSS DR7 data. Here only pixels surface brightness brighter than $30 \magarcsec$ are shown in the figure. The two  maps share the same color gradients as indicated by the colorbars. It is clear that the $PSF$ convolution produces a much smoother map and lots of discreet stellar light shown in the left panel disappear in the right panel (fainter than $30 \magarcsec$). This indicates that $PSF$ smoothing has a significant impact on the ICL calculation when applying a realistic surface brightness limit.  

After applying the above procedures, each grid of the mock image has a surface brightness. The  ICL fraction is  then defined as  the ratio of the total luminosity  of  all  grids  with  surface  brightness fainter than a given limit at $x$ band, $\mu_{x,limit}$, to the total luminosity in the galaxy group within the virial radius. It is also interesting to define another quantify as intra-cluster mass (ICM) fraction, which  is the  ratio of  all stellar mass in grids with surface brightness fainter than the  limit to the total stellar mass in the group. If  a constant stellar mass-to-light ratio is assumed, the two definitions will be equal. However, this assumption is not often valid and it is worth to check its influence on the estimation of ICL fraction. We will later see that the ICL fraction in term of light is lower than the stellar mass of the ICL component.

As galaxy luminosities vary with observational bands and the measured ICL fraction is usually made at a given band with a given magnitude limit, to compare the measured ICL fraction at different bands, it is important to know how to convert the SBL at different filters. As each galaxy has different star formation history and metallicity, the color is different for different galaxies. To illustrate this effect, we use a simple SSP evolution model with metallicity $Z = 0.02Z_\odot$ and a fixed stellar formation time at $z = 5$ to give the residual of surface brightness limit (relative to the V band magnitude limit), as a function of the effective wavelength of 10 different bands at different redshifts. 

\begin{figure}
\begin{center}
  \includegraphics[width=0.5\textwidth]{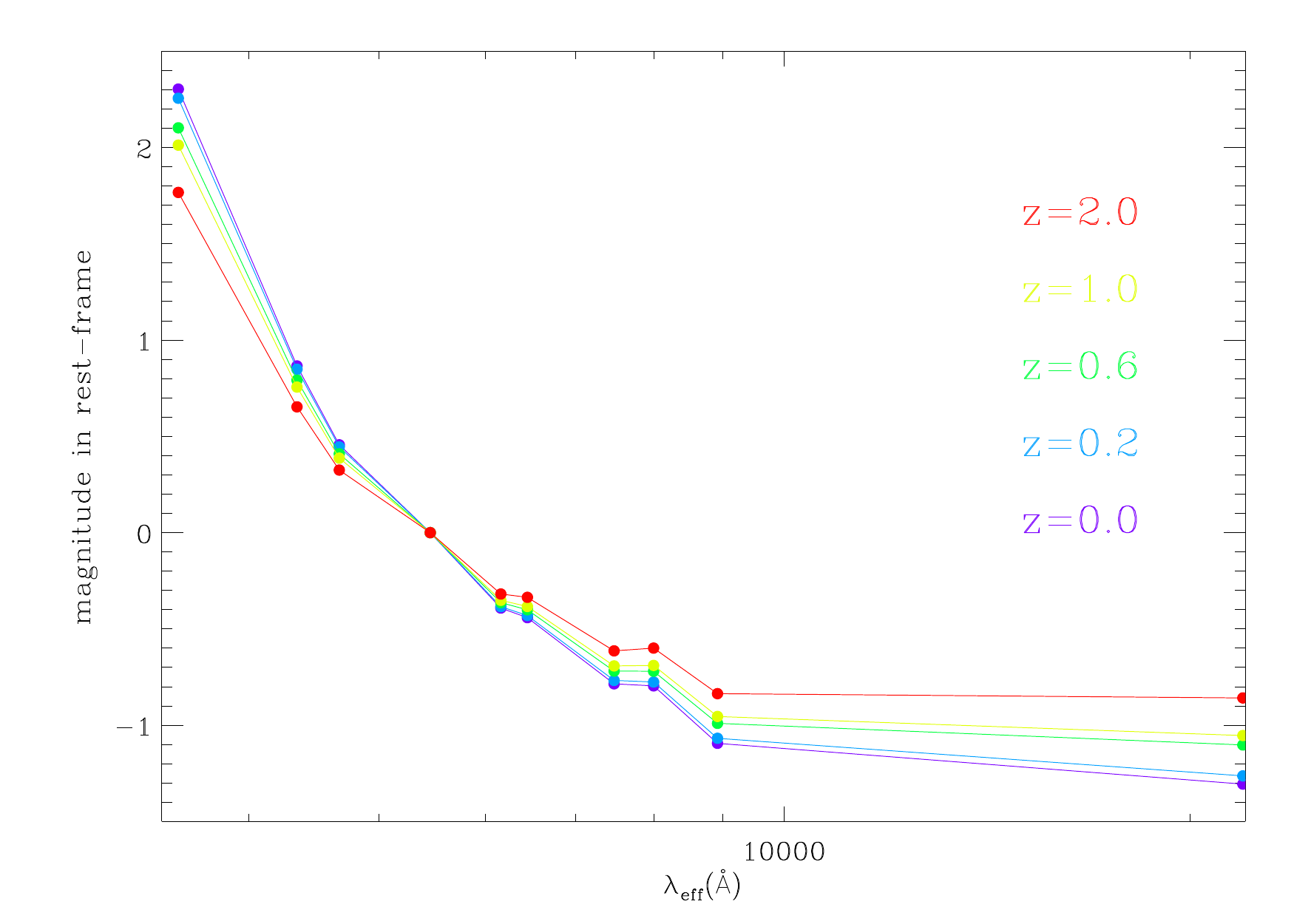}
    \caption{Given a simple SSP model, we obtain the magnitude (equivalent with surface brightness limit) for following ten bands at rest-frame, $u,B,g,V,r,R,i,I,z$, and $K$ bands from left to right, respectively. The abscissa gives the effective wavelength at each band, while the ordinate represents the equivalent surface brightness limit at these bands, normalized to the magnitude of $V$ band. Different redshifts are denoted by different colors.}
    \label{figure2}
\end{center}
\end{figure}

Figure~\ref{figure2} shows the dependence of brightness on observational bands and redshifts, where the $y-axis$ represents  the residual magnitude relative to the $V$ band magnitude. Different colors denote different redshifts. It is found that the residual is a function of wave length and redshift. After applying the conversion between different bands, the ICL fraction measured at different bands are expected to converge, as we will see below for most results. However, as the stellar population in each `galaxy' is not as simple as the model used here, we will see some difference which is expected. Note that the trend shown in Figure~\ref{figure2} is similar to that of Table 2 in \citet{BR07}.

\section{results}
\label{results}

\subsection{The impacts of pixel size and $PSF$ width on ICL fraction}
\label{influ_alpha_psf}

\begin{figure*}\epsscale{0.57}
  \plotone{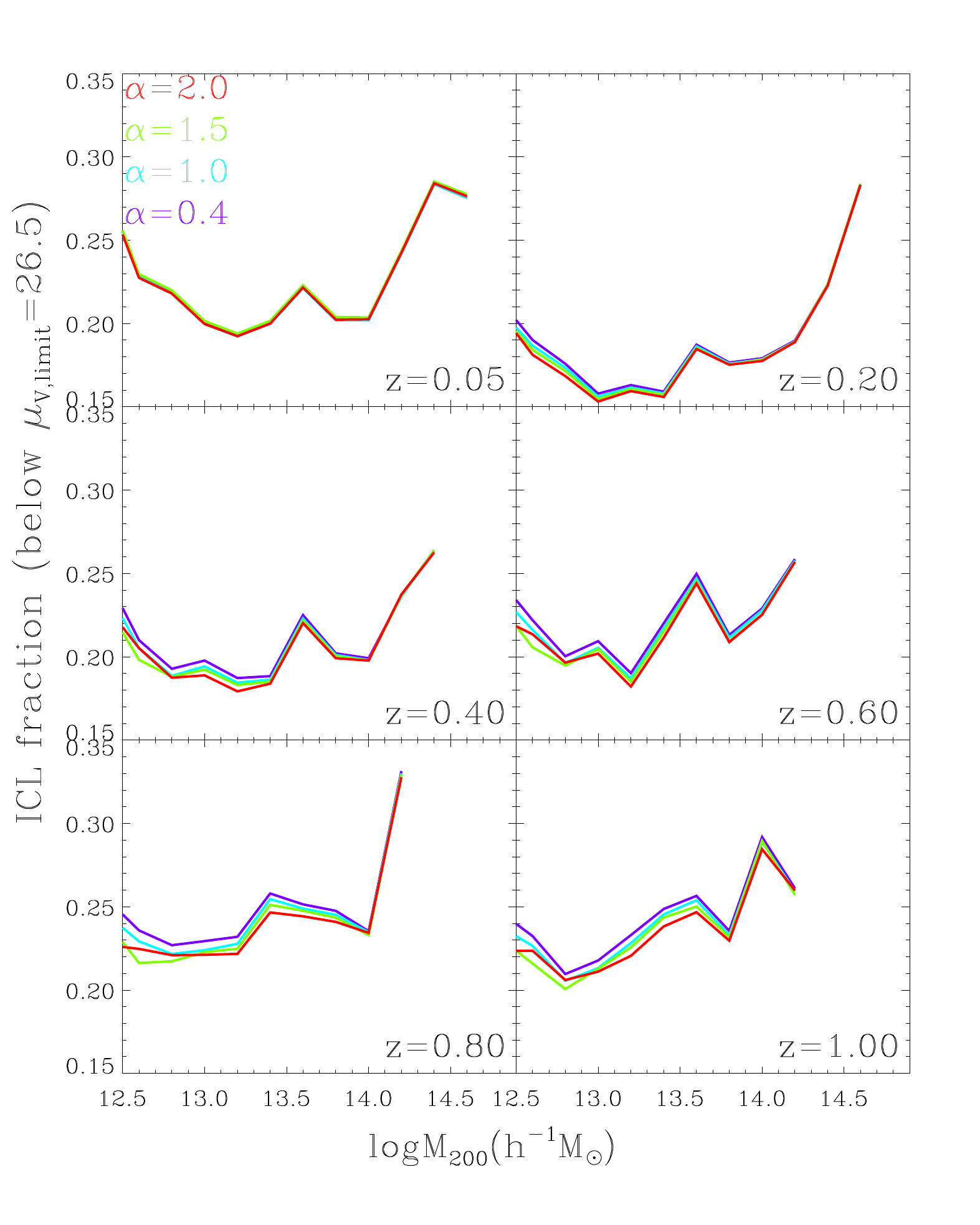}
  \plotone{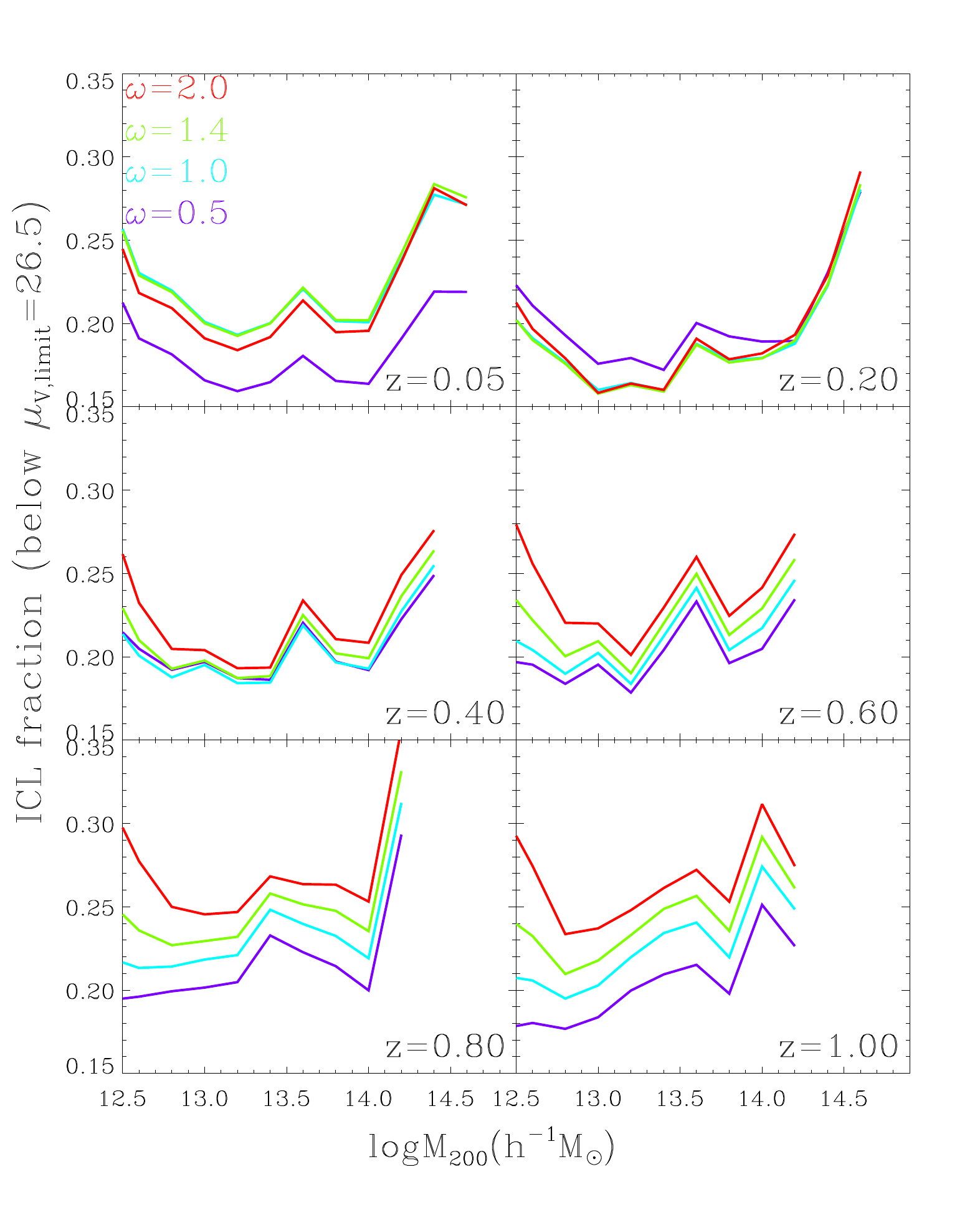}
    \caption{The measured ICL fraction for different angular pixel size $\alpha$ (left panels) and $PSF$ width $\omega$ (right panels) at $V$ band without taking into account the cosmological redshift dimming effect, namely omitting the redshift term in Equation~\ref{eq_lx}. The same SBL $\mu_{V,limit} = 26.5 \magarcsec$ is applied for all the results. For the results in the left or right panels, the same $\omega=1.43^{''}\sim1.4^{''}$ or $\alpha=0.396^{''}\sim0.4^{''}$ is adopted, respectively.}
    \label{figure3}
\end{figure*}

The impacts of angular pixel size $\alpha$ and $PSF$ width $\omega$ on the measured ICL fraction can be easily investigated using the mock images. In Figure~\ref{figure3} we show the measured ICL fraction as a function of halo mass for four different $\alpha$ (left panels) and four different $\omega$ (right panels) at a few redshifts. The results for different pixel sizes and $PSF$ widths are shown by the curves in different colors as indicated in the top left panels. Note that here we do not apply the cosmological redshift dimming effect (with neglect of the $(1+z)^{-4}$ factor in Equation~\ref{eq_lx}) and only the results at $V$ band with a fiducial SBL of $\mu_{V,limit} = 26.5\magarcsec$ are shown. We have tested that the effects of $\alpha$ and $\omega$ has a very weak dependence on the adopted bands and the choice of  SBL.

It is seen from the left panels of Figure~\ref{figure3} that the measured ICL fraction over all halo mass range and redshift shows almost no difference with the four adopted pixel sizes $\alpha  =  0.396^{''}\sim0.4^{''}$, $1.0^{''}$,  $1.5^{''}$, $2.0^{''}$. While in the right panel, larger $PSF$ width $\omega$  tends to give a higher ICL fraction, and this effect is independent of both halo mass and redshift. It  is also seen that the difference becomes smaller with bigger $\omega$.

Figure~\ref{figure3} shows that the influence of $\omega$ is much more obvious than that of $\alpha$. As surface brightness is in unit of area, it is not  surprising that pixel size has less  influence on the ICL fraction for the highly clustered region because more particles will be assigned to the grid when the pixel size becomes larger, and eventually keep the surface brightness unchanged.  On the other hand, $\omega$ controls the width of the convolution kernel, i.e., the smoothing level of the image. Larger $\omega$  will produce smoother  image, resulting in more pixels with lower surface brightness \footnote{Note that the angular pixel size of a modern {\small CCD} is much smaller than the $PSF$ width.}. This smoothing effect is  more significant for diffuse or under-dense regions, such as low mass halos and the halos at high redshifts.

\subsection{The impact of surface brightness limits on ICL fraction}
\label{influ_uv}

Surface brightness  limit  is the simplest and widely used method for identifying ICL fraction. It is important to investigate its capability and limitations. Since the effects of pixel size and $PSF$ width have been known from Figure~\ref{figure3}, we fix $\alpha  = 0.396^{''}$ and $\omega=1.43^{''}$  in the following analysis and the cosmological redshift dimming is also included in this section. 

Apparently, the magnitude limit and observational band are the main factors in the SBL method. In a few studies on observations and semi-analytical models, a faint surface brightness limit $\mu_{V,limit}=26.5\magarcsec$ is used to separate ICL from the galaxies \citep[e.g.,][]{Vilchez-Gomez1994, Zibetti08, Rudick06a, Rudick11, Feldmeier04a}. \cite{Cui2014a} used a slight brighter magnitude limit with $\mu_{V,limit}=23.0$, $24.7\magarcsec$ to make comparison between a dynamical method and surface brightness limit method.  However, it  is not clear how the results will be affected by these adopted values. As a references to these studies, in this paper we adopt three SBLs with $\mu_{V, limit}  =23.0$, $24.7$, $26.5\magarcsec$  at rest-frame to distinguish BCG and ICL component. For identifying ICL at other bands, we use the conversion factor given in Figure~\ref{figure2}. Again we note that the conversion factor is from a very simple SSP model and it is not surprising that the measured ICL fractions at different bands will have slight difference.

\begin{figure*}
\begin{center}
   \includegraphics[width=0.95\textwidth,height=0.65\textheight]{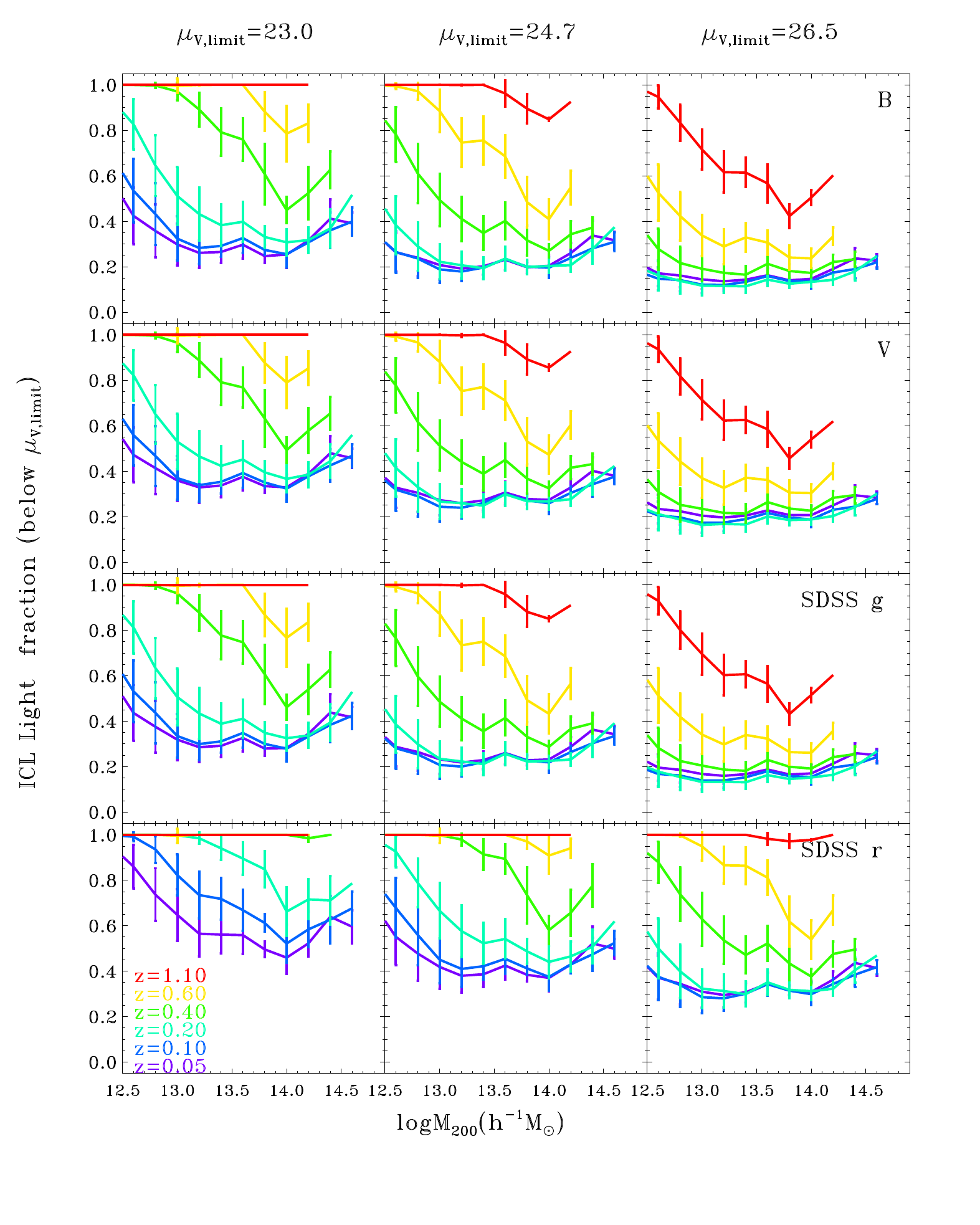}
    \caption{The measured ICL fractions as a function of halo mass $M_{200}$. Different columns use different SBLs as indicated on title of the top row. Note that these showed SBLs are at $V$-band, and SBLs for other bands have been converted in order to make fair comparison. Each row shows the results at different observational bands, $B$, $V$, SDSS $g$ and SDSS $r$ from top to bottom respectively. The surface brightness is calculated with $(1+z)^{-4}$ in term of $I_{x}$. Here, the same $\alpha=0.396^{''}$ and $\omega=1.43^{''}$ are adopted. As indicated by the legend in the bottom left panel, different line colors represent different redshifts. The error bars are calculated by $\sigma=\sqrt{\sum{\frac{(x_{i}-\bar{x})^2}{n}}}$.}
    \label{figure4}
\end{center}
\end{figure*}

Figure~\ref{figure4} shows the results of the measured ICL fraction as a function of halo mass at different redshifts. This figure contains two distinct features. Firstly, by comparing different columns, it is not surprising to find that ICL fraction increases with  brighter SBL. This increase is weakly dependent on both halo mass and redshift, and it is very similar at all bands. This increase of ICL fraction is simply caused by the fact that when a brighter limit is applied, more stellar particles will be assigned to ICL. Secondly,  groups at higher redshifts tend to have a higher ICL fraction for a given SBL. This result is mainly caused by the cosmological redshift dimming (Equation~\ref{eq_ux}) and partly by the `galaxy' evolution itself, as high-redshift `galaxy' is still forming with less stars (and thus fainter) compared to its counterpart at lower redshifts. Therefore, it is not surprising to find that the measured ICL fraction reaches 100 percentages at high redshift with a brighter (not realistic) SBL. 

With faint SBLs, e.g., $\mu_{V,limit}=26.5 \magarcsec$, it is interesting to find that the ICL fraction does not depend on halo mass at low redshifts. This can be explained as following. The stellar distribution of all halos at low redshift is similar and most of  the stellar particles in very faint (outer) region of halos have been counted as ICL component (see Figure~\ref{figure7}). More importantly, the observational influence, i.e., the cosmological redshift dimming and $PSF$ effect, can be ignored because of the high stellar density within halos at low redshift, as can be demonstrated later in Figure~\ref{figure6}. These facts cause the similar ICL fraction for all halos at low redshift with faint SBL.

Figure~\ref{figure4} shows a major limitation of the SBL method to identify ICL that a very faint magnitude limit is needed otherwise all stars could be identify as ICL. It is also seen from the figure that a magnitude limit at $V$ band fainter than $26.5$ is needed to identify the ICL for group at $z=1$.

By using the simple SBL conversion in Figure~\ref{figure2}, we expect to obtain approximately similar ICL fraction at other bands. However, the results for $r$ band seem to have a higher fraction than others. This indicates that the $r$ band magnitude limit is brighter than that obtained from the conversion. One possible reason is that the simulated galaxies have a more extend star formation history and lower metallicity, thus the $V-r$ color is bluer than that obtained from a simple SSP model with star formation epoch at $z=5$ with metallicity of $Z=0.02Z_{\odot}$.

\begin{figure*}
\begin{center}
  \includegraphics[width=0.95\textwidth,height=0.65\textheight]{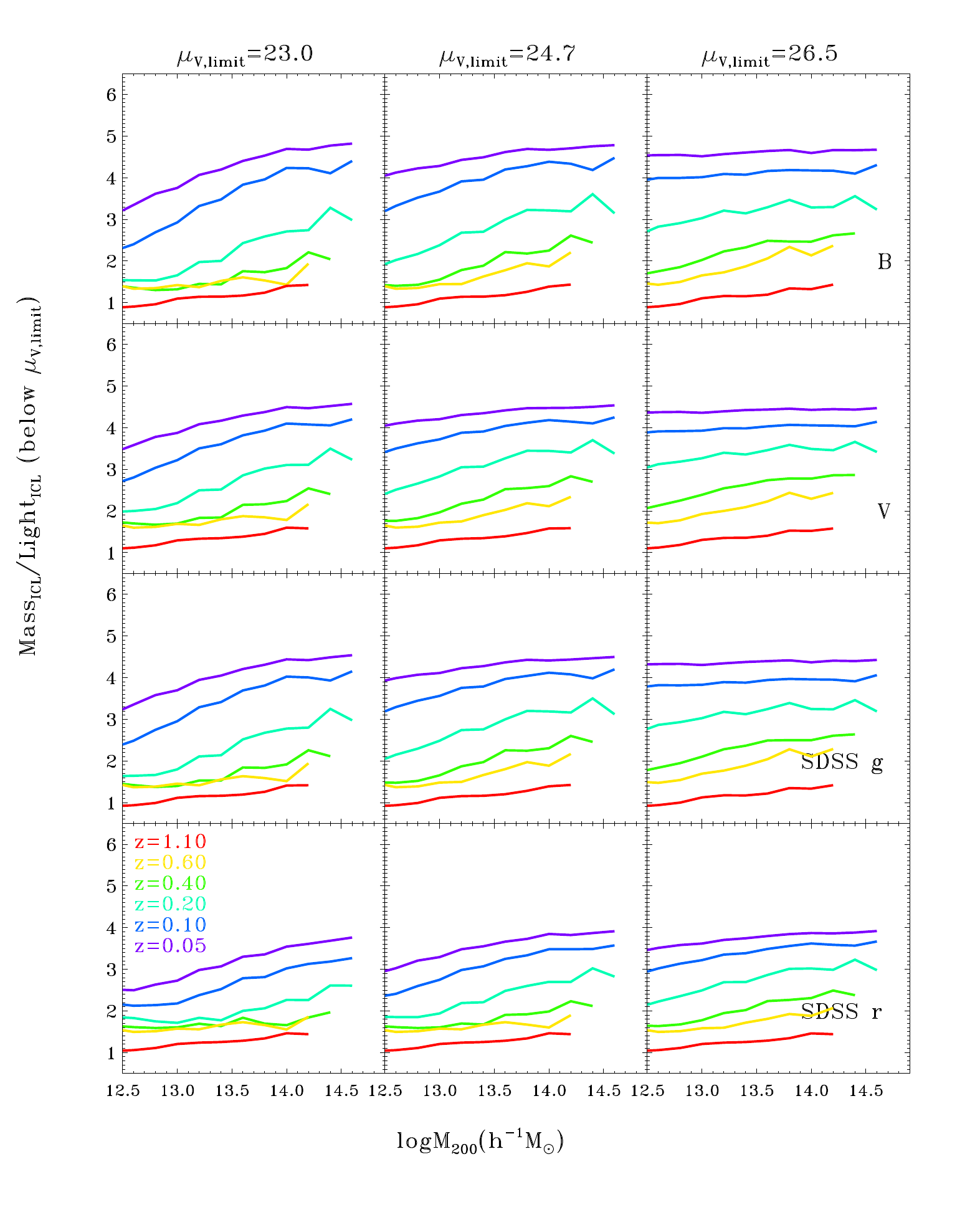}
    \caption{Same as Figure~\ref{figure4}, but for the mass-to-light ratio of ICL
    component.}
    \label{figure5}
\end{center}
\end{figure*}

In Figure~\ref{figure5}, we show the ratio between the mass fraction locked in the ICL component to the light fraction in the ICL component.  It is found that ICL fraction estimated from luminosity is systematically lower than that estimated from the stellar mass. The difference is slightly bigger for fainter SBL and shows an obviously dependence on the redshift and halo mass. These differences indicate that there is more stellar mass locked in ICL component due to the higher mass-to-light ratio of ICL than that of the main `galaxy' within the halo. The higher mass-to-light ratio of ICL indicates that color of stars in ICL component is redder than the main `galaxy'. We have checked the luminosity-weighted metallicities and ages of star particles assigned to the `galaxies' and ICL component, and found that the ICL stars have higher metallicity and older age than `galaxies'. In this paper we do not investigate in detail which factor contributes mostly to the higher mass-to-light ratio of ICL component, and such analysis would be useful in future work for comparison with observations.

\begin{figure*}
\begin{center}
  \includegraphics[width=0.3\textwidth,height=0.65\textheight]{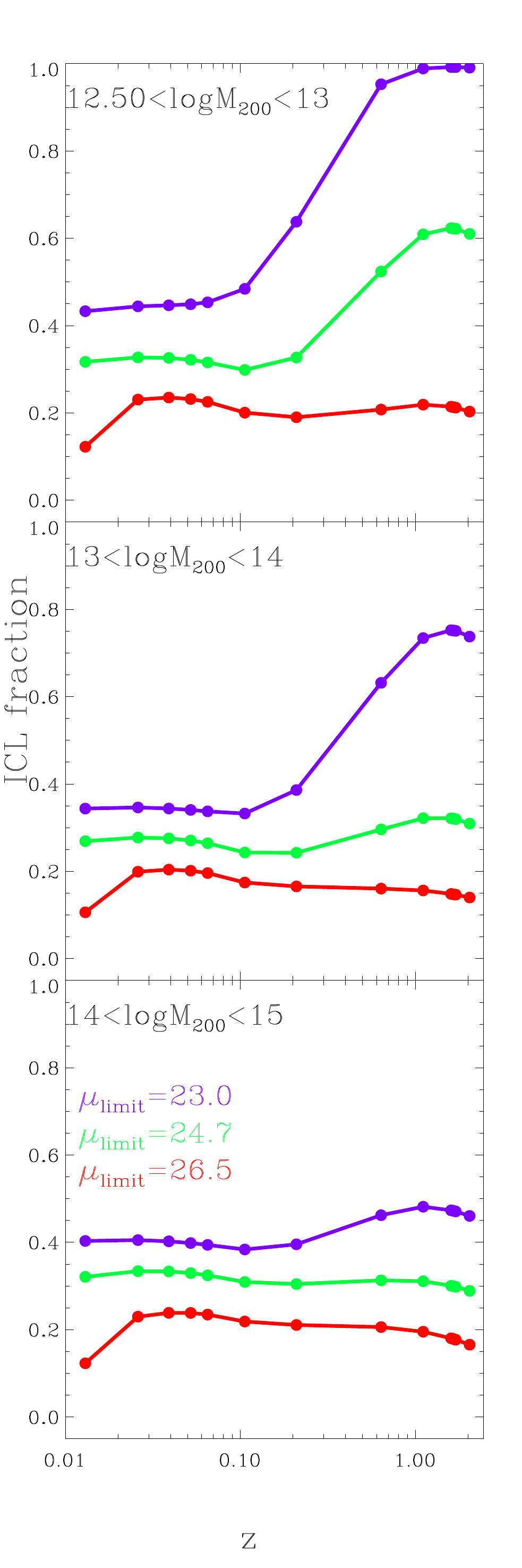}
  \includegraphics[width=0.3\textwidth,height=0.65\textheight]{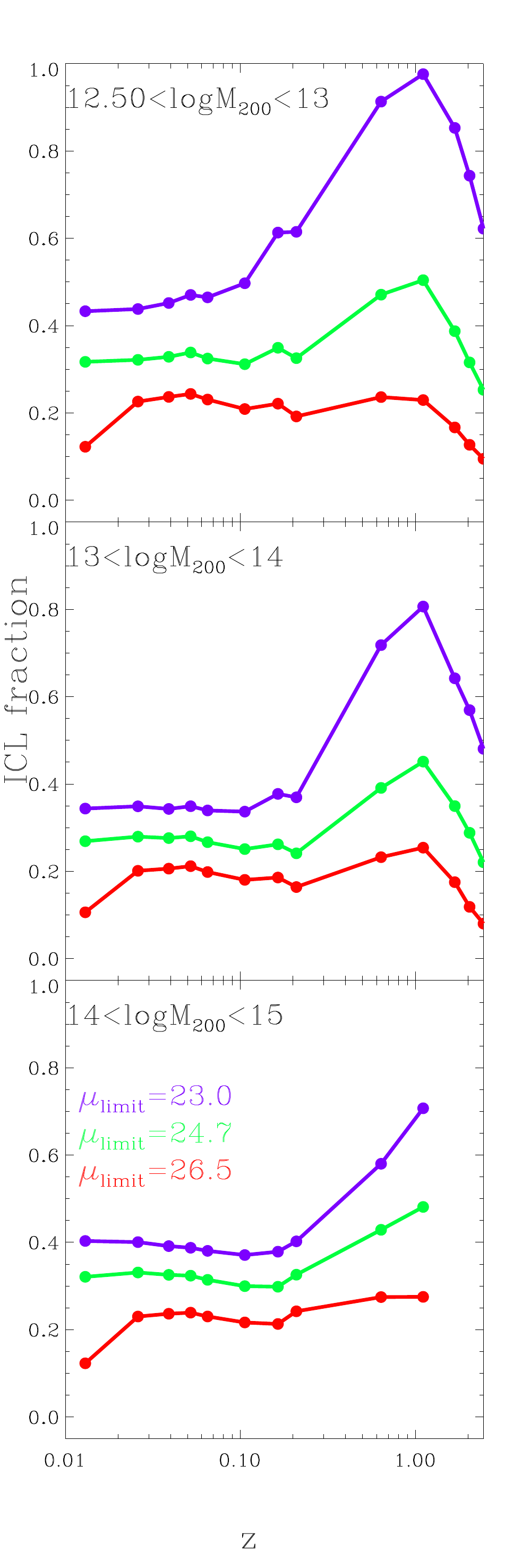}
  \includegraphics[width=0.3\textwidth,height=0.65\textheight]{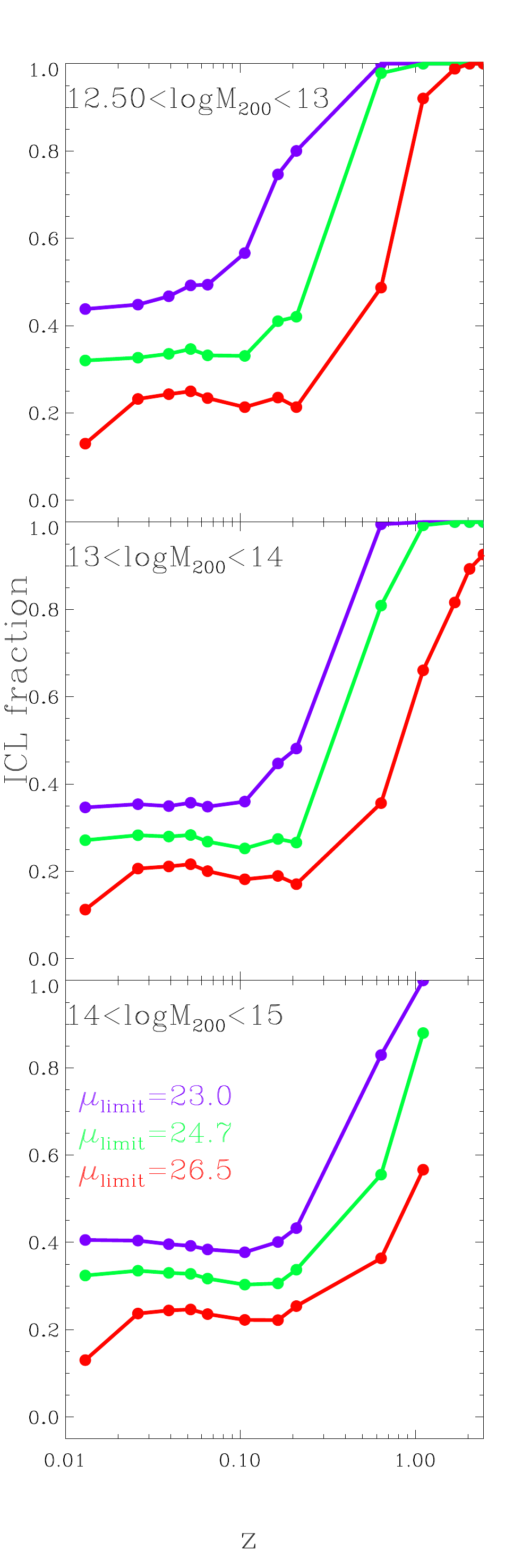}
    \caption{The $V$-band measured ICL fraction as a function of redshift at different mass bins. The left panel shows the ICL fraction by putting $z = 0$ `galaxies' to high redshifts. The middle panel is the result from the simulated galaxies at corresponding redshifts. The cosmological redshift dimming is not taken into account for the results in these two panels. The right panel is similar to the middle panel, but taking into account the cosmological redshift dimming. The same pixel size and $PSF$ width are adopted as in Figure~\ref{figure4}. The lines in different colors represent results for different SBLs.}
    \label{figure6}
\end{center}
\end{figure*}

In order to separate the intrinsic 'galaxy' evolution from the observational cosmological redshift dimming effects on ICL fraction, we further compare  the measured ICL fraction evolution as a function of redshift in Figure~\ref{figure6}. In the left panel, we only use the `galaxies' from $z \sim 0$ snapshot and shift them to different redshifts to produce mock images. In the middle panel, we use `galaxies' produced in the simulation at different redshifts. In these two panels we do not apply the cosmological redshift dimming effect (omitting the $(1+z)^{-4}$ term in Equation~\ref{eq_ux}). The right panel shows the results using the `galaxies' produced in the simulation at different redshifts and with cosmological redshift dimming included. Different rows show the results in different halo mass bins, and the lines in different colors  indicate different SBLs at $V$ band. Here results for more redshift outputs are plotted than those in Figure~\ref{figure3} and Figure~\ref{figure4} to obtain smoother curves.

This figure shows some interesting results. Firstly, the left panel represents the pure effects of $PSF$ and cosmological geometry on ICL fraction. It is seen that the ICL fraction is almost flat across all redshifts, except for the results with the highest SBL. By simply putting $z = 0$ `galaxies' to higher redshifts, regardless of cosmological redshift dimming, only the pixel size and the $PSF$ can have effects on the ICL fraction. Since the pixel size has no significant effect (see Section~\ref{influ_alpha_psf} for details), the redshift evolution of ICL fraction showing in the left panel of Figure~\ref{figure6} should only lie in the $PSF$ effect, which is much more significant for fainter regions. Therefore, we are expecting a slightly larger increase of ICL fraction for the halos with smaller mass, and more obvious evolution of ICL fraction with brighter SBL. 

Furthermore, given a faint SBL, i.e., $\mu_{V,limit}=26.5\magarcsec$, there is a flat redshift evolution, as can be seen from both the left and the middle panel. The redshift evolution is also flat for all halos over the redshift range of 0 and 0.2, even taking into account the cosmological redshift dimming effect, as shown in the right panel. 

In the left panel of Figure~\ref{figure6}, the curves slightly bend down at high redshift. This is from the cosmological geometry effect. It has been given that, $D=\alpha*\frac{ a_0 r }{ 1 + z } = \frac{ c\alpha }{ H_{ 0 }(1+z) }\int_{ 0 }^{ z }{ \frac{ dz }{( \Omega_{ \Lambda,0 } + \Omega_{ m,0 }( 1+z )^3 )^{1 / 2} } }$. At a fixed {\small CCD} pixel size, $D$ increases with redshift until $z\sim1.6$, and then decrease. $PSF$ can make the target smoother and fainter when $D$ is larger. Therefore, if cosmological redshift dimming is not taken into account, the target moving to high redshift beyond $z\sim 1.6$ will become fainter and then brighter, because its image will become smoother and then turn sharper. In such a case, from the peak redshift to high redshift, ICL fraction will decrease. The curves also have dependence on the SBLs and halo mass, as their peaks become more prominent with a higher SBL or for less massive halos. One can speculate that the $PSF$ has different effect at inner and outer region of BCG, in good agreement with the $PSF$ influence, which has more significant effect on fainter components.

Secondly, as shown in the middle panel, the measured ICL fraction is higher for less massive halos and demonstrates a strong redshift evolution. Comparing to the left panel, ICL fraction shows a deeper drop at higher redshifts, especially for less massive halos. This could reflect the facts that `galaxies' at higher redshift are less massive and more compact than those directly shifted from $z\sim0$ snapshot, while the diffuse stellar components are consecutively produced. Those curves show a peak at $z\sim1$, later than those in the left panel. These two facts regulate the redshift evolution of ICL fraction. 

Finally, with the cosmological redshift dimming, for all halos within mass bins, there is a significant redshift evolution of ICL fraction that increases sharply with redshift. With the brightest SBL adopted here, the measured ICL fraction will quickly reach 100 percentages at redshifts beyond $z > 0.6$ and $z >1$ for less massive halos and massive halos respectively, as can be seen from the right panels.

\subsection{The evolution of ICL profile}
\label{Sec_profile}

\begin{figure} \epsscale{1.1}
  \plotone{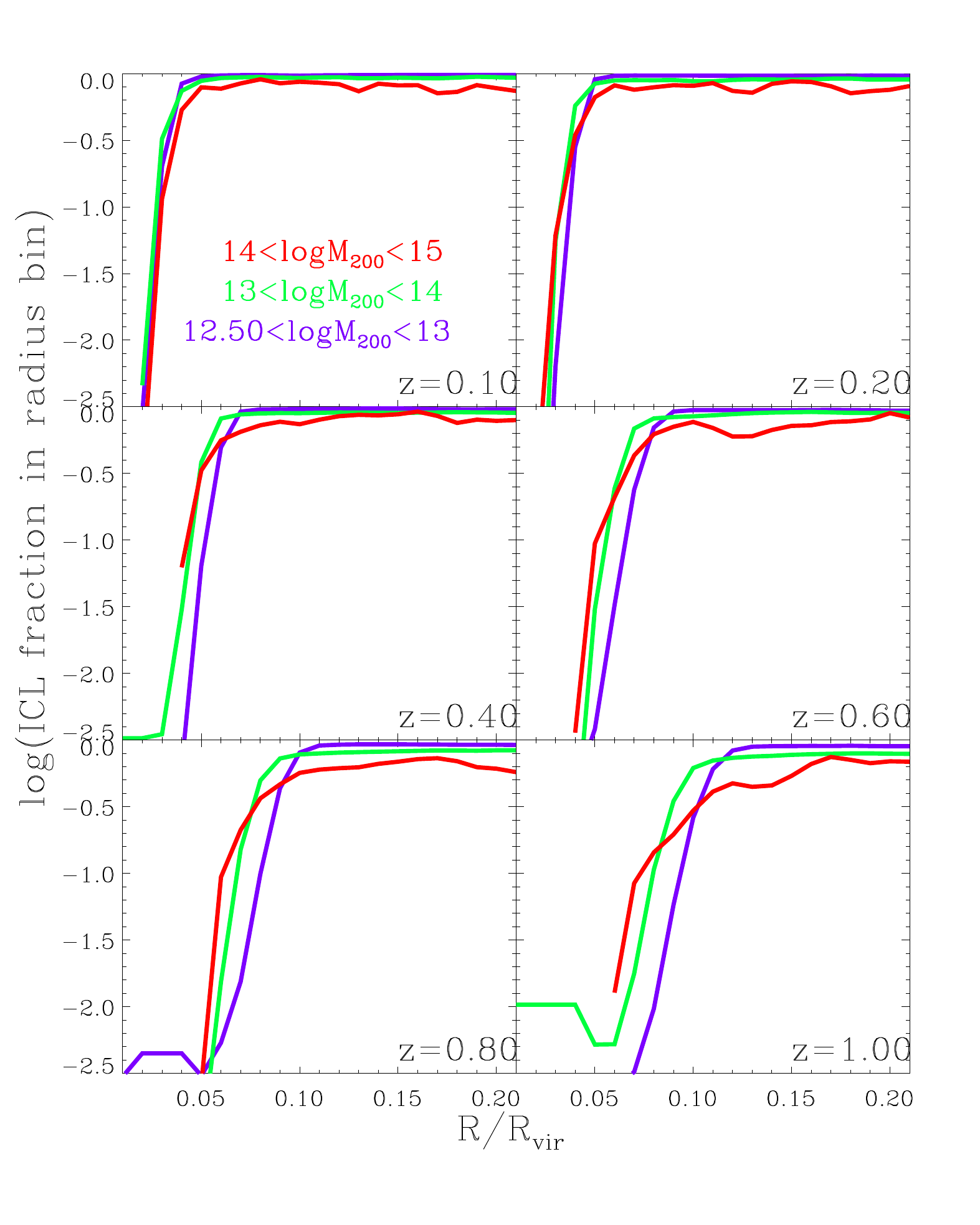}
     \caption{The measured ICL profile at different redshifts (different panels) and different halo mass bins (different colors). Here $\alpha=0.396^{''}, \omega=1.43^{''}$, $\mu_{V,limit}=26.5\magarcsec$ are adopted and there is no cosmological redshift dimming effect. The y-axis represents the value of ICL luminosity divided by total luminosity at the radius bin $R\sim R+dR$.}
     \label{figure7}
\end{figure}

Another interesting problem is the radial distribution of ICL component around BCG. If the ICL is the byproduct of galaxy formation, its spatial distribution will be expected to give information of the cluster formation at current stage. We calculate the abundance of ICL fraction at each radius bin normalized by the halo virial radius $R_{vir}$. Figure~\ref{figure7} shows the radial profile of ICL fraction at a few redshifts. The profiles seem to be independent of halo mass, but slightly extend with increasing redshift. It is interesting to see that almost all the star particles distributed outside of a certain distance are counted as ICL component. As most of the ICL stars are assumed to be formed through galaxy merger or stripping, this characteristic radius can indicate an important range of galaxy merger events.

This normalized characteristic radius, where the measured ICL fraction reaches 100 percentages, slightly increases from $\sim 0.04 R_{vir}$ at low redshift to $\sim 0.1 R_{vir}$ at high redshift. This change could be caused by the increasing of the halo virial radius towards $z \sim0$ but with a fixed physical radius where ICL fraction $\sim 1$, or the physical characteristic radius decreases with redshift. Using several randomly selected halos to check both the evolution of halo virial radius and the physical characteristic radius of ICL, we find that the main evolution is the decrease of the physical radius of ICL.

\begin{figure*} \epsscale{1.1}
  \plotone{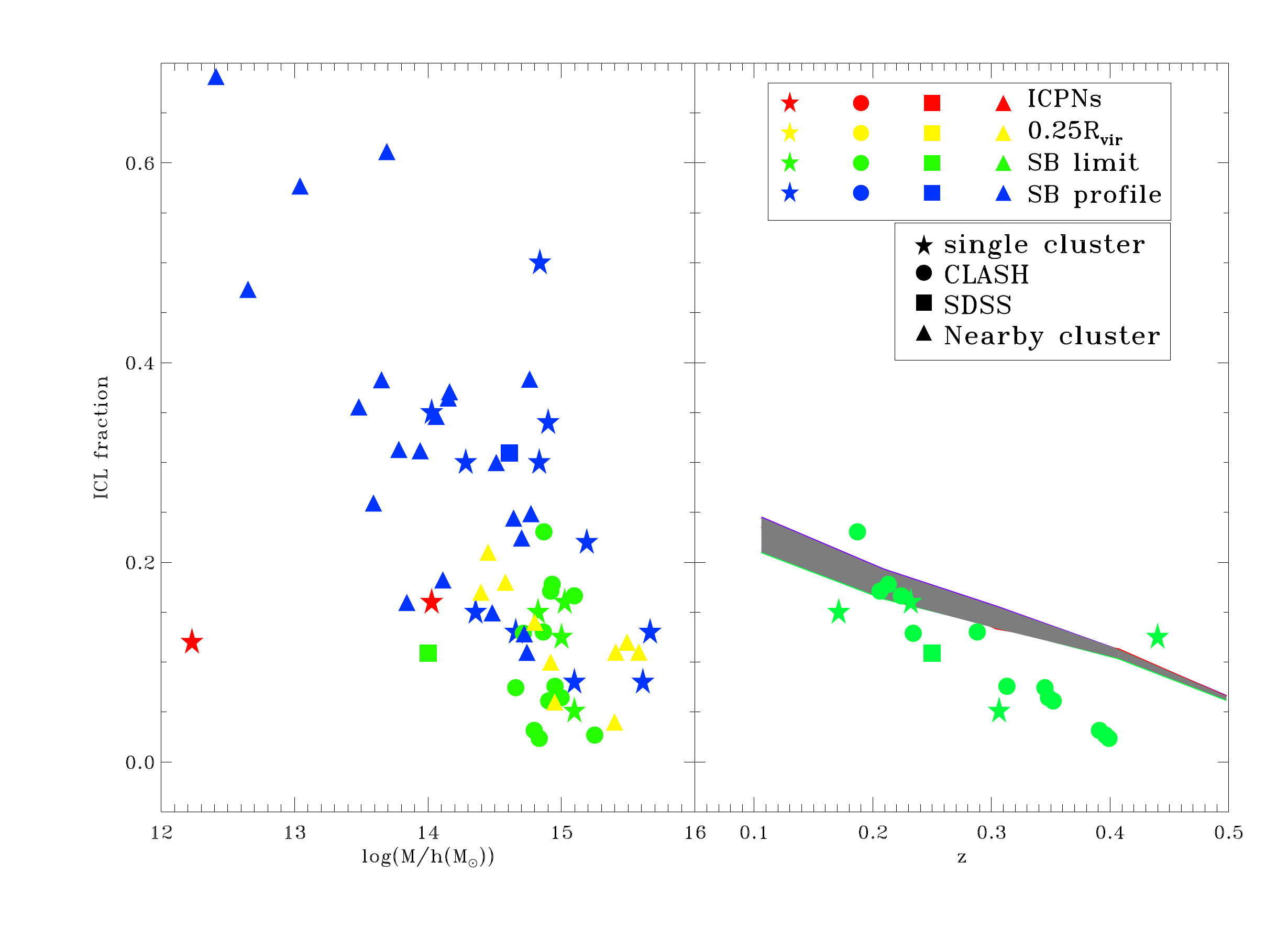}
     \caption{Comparison between the observed ICL fractions as a function of halo mass (left panel) and redshift (right panel). The denotation of symbols is as following. The five-pointed stars, circles, squares, and triangles represent the data from single clusters$^{1\sim15}$, CLASH$^{16}$, SDSS$^{17,18}$, and nearby clusters$^{19\sim21}$, respectively. In the right panel, we only plot the observational results defined by the surface brightness limit method, as the same method adopted in our mock observations. And, the gray shaded region is our result at $12.5< \log M_{200} (\msunh)<15$. The value of SBL used is $\sim24.0\magarcsec$ in V band, which is similar with $\mu_{B}=25.0\magarcsec$ in \cite{Burke15}, applying the conversion relation shown in Figure~\ref{figure2}. We correct the SBL with $2.5\log(1+z)^4$ for our simulated data to make a fair comparison with the data of \cite{Burke15} in the right panel. Width of $PSF$ and pixel size are those used in CLASH data, $0.75^{''}$ and $0.05^{''}$ respectively.
\vspace{6pt}\\
1: \cite{Longobardi15}, 2: \cite{MT14}, 3: \cite{Presotto14}, 4: \cite{Jee10}, 5: \cite{Vilchez-Gomez1994}, 6: \cite{Uson1991}, 7: \cite{Bernstein1995}, 8: \cite{TF1995}, 9: \cite{Feldmeier04b}, 10: \cite{Feldmeier02}, 11: \cite{Gonzalez05}, 12: \cite{Tyson1998}, 13: \cite{SK1994}, 14: \cite{Feldmeier04c}, 
15: \cite{Feldmeier04a}, 16: \cite{Burke15}, 17: \cite{Budzynski14},18: \cite{Zibetti08}, 19: \cite{Gonzalez07}, 20: \cite{Gonzalez05}, 21: \cite{KB07}. }
     \label{figure8}
\end{figure*}
\section{Observational and theoretical constraints on the ICL fraction}
\label{comparison}

In Table~\ref{tab:fit}, we compile data from both observations and theoretical predictions. The observational data can be roughly separated into four categories using the methods described in Section~\ref{sec_intro}. The individual clusters are categorized as $single \ cluster$ in our comparison. Table~\ref{tab:fit} provides only the mean parameters for the CLASH and Nearby Cluster populations, while the accurate ICL fraction of each cluster is plotted in Figure~\ref{figure8}. For various theoretical results, the physical models applied and the methods to identify ICL are quite diverse. Therefore, as we have mentioned, it is not capable to make a fair comparison if the definition of ICL is obviously not similar to that of observation. This is the reason why we compile only the observational results and our predictions in Figure~\ref{figure8} that illustrates the ICL fraction as a function of halo virial mass and its redshift evolution. 

In Figure~\ref{figure8}, the single clusters are represented by solid stars, which can be separated into three sets according to their ICL identifications. The first set includes three data points (red solid stars) \citep[]{Longobardi15, Feldmeier04b, Feldmeier04c} defined by {ICPNs} (or GCs) method because they are very low redshift local targets. The second set includes the results for other five clusters (green solid stars) \citep[e.g.,][]{MT14,  Presotto14} obtained using the SBL method. Note that their SBLs are different with each other, and the higher ICL fraction is caused by the brighter SBL, except that A2390 has a brighter SBL in $B$ band \citep[]{Vilchez-Gomez1994},  but shows a lower ICL fraction. The discrepancy is caused by the brighter optical detection depth in \cite{Vilchez-Gomez1994}. Generally, the surface brightness of ICL component is so faint that a optical depth should be reached to detect such components. And the last set includes the remaining single clusters (blue solid stars) \citep[e.g.,][]{Uson1991, SK1994} obtained by the SB profile method. After checking their SB profiles, it is found that the brightness profiles excess the $r^{1/4}$ law cause the high observed ICL fractions. In \citet{Gonzalez00} the ICL fraction of A1651 is found to be $<\sim2\%$ . With more accurate calculation, its ICL fraction becomes $\sim13\%$ (see Table 4 in \cite{Gonzalez05} and Table 1 in \cite{Gonzalez07}).

The SDSS clusters can be separated into two types by their ICL definitions. ICL for a fraction of clusters (blue square) is defined by a modified SB profile method, and the mean fraction is 0.31. The others (green square) are obtained by the SBL method with a value of $25.0 \magarcsec$ in $r$ band, and the mean fraction is $\sim15\%$. Note that the pixel size $0.396^{''}$ and $PSF$ value $1.43^{''}$ applied in our calculation are adopted from SDSS. 

\cite{Burke15} uses the SBL method to define ICL for all the CLASH clusters (green solid circles) with $\mu_{limit}=25\magarcsec$ in the rest-frame $B$ band,  where the image pixel size and $PSF$ width are $0.05^{''}$ and $0.07^{''}\sim0.15{''}$ \citep[e.g.,][]{Zitrin2011,Postman2012,Merten2015}, respectively, much smaller than those of SDSS. As discussed in Section~\ref{influ_alpha_psf} for the $PSF$ smoothing effect, the smaller PSF width for CLASH data may partly account for the lower ICL fraction than SDSS results, apart from the dependence on halo mass. In addition, the SBLs vary for different clusters, and thus the calculated ICL fraction shows a big scatter over a narrow halo mass range.

For the nearby clusters, two methods are used to define ICL. For the first one, ICL (blue triangle) is estimated through a modified SB profile method \citep[]{Gonzalez05, Gonzalez07}, with  $\alpha=0.7$, $\omega=1.4$ , over a redshift range of $0.03<z<0.13$. While for the other one, ICL (yellow triangle) is defined as the flux at the region outside of $0.25R_{virial}$ \citep[]{KB07}, with $\alpha=0.345$, $\omega\sim1.0$, over a redshift range of  $0.05<z<0.3$. It is seen that the former obtained a much higher ICL fraction and a bigger scatter. We have checked their surface brightness profiles, and found that \cite{Gonzalez05, Gonzalez07} split the BCG from ICL at smaller radius than that in \cite{KB07}, so the SBL in Gonzalez et al. is actually higher than the latter, leading to a higher ICL fraction.

In \cite{Longobardi15}, their ICL is defined by the ICPNs method (red symbol), and according to their definition, the ICL fraction is obtained as intra-cluster PNs  (ICPNs) divided by the total PNs (IC PNs + BCG PNs), namely the ratio of ICL/(BCG+ICL) in luminosity. Therefore, their derived ICL fraction is different from others as they did not consider the contribution from other galaxies than the central BCG. Thus the ICL fraction serves as only an upper limit. Note that the significant scatter in \cite{Longobardi15} is caused by the small number statistics of ICPNs data.

It is also found that the observed ICL fraction obtained by the SB profile method (blue symbol) varies within a large range. The SB profile method normally implies a bright SBL value for halos with low and intermediate mass, and thus causes a larger ICL fraction. Actually, the ICL fraction calculated by this method is more close to ICL/(ICL+BCG). The observational data defined by the SBL method (green symbol) locate in a smaller region than the SB profile data. 

For the data defined by the $0.25R_{vir}$  method (yellow symbol), according to what we have discussed in Section~\ref{Sec_profile}, the characteristic radius evolves with redshift. Thus the constant radii for halos at different redshifts should no longer be valid to accurately define ICL, so as to cause an increasing observed ICL fraction with mass. The ICL fraction of all the data in the left of Figure~\ref{figure8} apparently shows an obviously dependence on the halo mass, which differ from ours results.

In the right panel of Figure~\ref{figure8} we show the redshift evolution of the ICL fraction for a few observational results and our simulation predictions (shown in the gray shaded region), where the ICL is defined by the SBL method. Note that both observational and theoretical results have included a surface brightness limits which evolves with redshift as $2.5\log(1+z)^{4}$ \citep[]{Burke15} and other similar observational parameters to make a fair comparison. For the CLASH results, the ICL fractions drop from 0.23 to 0.02 with redshift changing from 0.18 to 0.4. We found that the decrease of ICL fraction with redshift is mainly driven by the evolution of surface brightness limit. The general trend in observations agrees with our predictions although the observational dependence on redshift is slightly stronger.

\section{Discussion and Summary}
\label{summary}
 
In this paper, we make mock observational images of galaxy groups and clusters from a cosmological hydro-dynamical simulation and investigate the  measured ICL fraction identified by the SBL method. These mock images are smoothed by a gaussian kernel with $PSF$ width $\omega$ over all image pixels with size $\alpha$. Results at four different filter bands $B$, $V$, SDSS $g$ and SDSS $r$ are presented in this paper. The main results can be summarized as follows.

\begin{enumerate}
    \item $PSF$ width $\omega$ has a clear effect on the measured ICL fraction, while pixel size $\alpha$ has little effect. Larger $\omega$ leads to a strong smoothing that leads to a higher ICL fraction. The cosmological redshift dimming and observational detection depth (or SBL) have significant impacts on the measurement of ICL component. 

    \item As shown in Figure~\ref{figure4}, the  measured ICL fraction strongly changes with the surface magnitude limit so that ICL fraction decreases with fainter SBL, ranging between 0.2 and 0.4 for massive halos at low redshift. The  measured ICL fraction also depends mildly on halo mass. With a certain SBL and a fixed redshift,  the measured ICL fraction decreases with increasing halo mass and becomes almost independence on halo mass beyond $M_{200} \gtrsim 10^{13} \msunh$, especially at low redshift. In particular, with a faint SBL, i.e., $\mu_{V,limit}=26.5 \magarcsec$, the measured ICL fraction does not depend on halo mass at low redshifts. There is almost no difference among $B$, $V$, SDSS $g$ band, while the results for SDSS $r$ may be largely affected by the conversion factor derived from a simple SSP model.

    \item Using this SBL method, the measured ICL fraction clearly shows apparent redshift evolution. It dramatically increases with redshift, due to the efficient cosmological redshift dimming effect, as can be seen from Figure~\ref{figure4} and the right panel of Figure~\ref{figure6}. Removing the effect of cosmological redshift dimming, as shown in the middle panel of Figure~\ref{figure6}, the measured ICL fraction increases with redshift, reaches a peak around $z \sim1$ and then decreases. 
    
     \item Figure~\ref{figure5} shows that mass and luminosity fraction of intra-cluster stellar component are different and there is more mass locked in the ICL component than the light. This is due to the larger mass-to-light ratio in ICL component than the main `galaxy'. Our results indicate that the use of a constant mass-to-light ratio at high surface brightness levels will lead to an underestimate of the mass in ICL.
     
    \item Figure~\ref{figure8} shows that current measurement of ICL from observations has a large dispersion due to different ICL definitions and observational environment. The general trend of ICL redshift evolution in observational results of \cite{Burke15} agrees with our theoretical predictions, using same ICL definition and similar surface brightness limit, $PSF$ and pixel size. Given the above, we emphasize the importance of using the same method and observational parameters (e.g., surface magnitude limit, $PSF$ and pixel size) when observational results are compared with the theoretical predictions.
     
\end{enumerate}

\vspace{10pt}

Although the used hydro-dynamical simulation in this work is not perfect, and may have influence on the results, we claim that our qualitative conclusions are solid, especially the effects of SBL, $PSF$ and cosmological redshift dimming. The results for low mass halos and halos at high redshift are not reliable due to the lower mass resolution of the simulation. In order to fully understand the physics of ICL component, more high-resolution simulations with proper baryonic physics included are required. For example, including AGN feedback in simulations will give a large change on ICL fraction \citep[]{Cui2014a}. 

The SBL method may also help to improve the studies of galaxies. In some simulations, the predicted stellar mass functions of galaxies are roughly consistent with observational data in the range of intermediate stellar mass, while the low-mass end is over-predicted and the high-mass end is over- or under-predicted \citep[e.g.,][]{Liu10}. \cite{Liu10} also checked the mass function for different type galaxies, and found that the identification of substructures (SUBFIND) in theories assigns too many stars, which causes that the results are not consistent with observations. The over-assigned stars must come from the regions around galaxies. If these stars are re-assigned to ICL component, the stellar mass function may decrease to match with data. Furthermore, in observation, very faint galaxies might be omitted, or confused with ICL due to their faint surface brightness. In addition, if there is a large fraction of ICL component which has been omitted from observations of galaxy groups, the so-called missing baryon problem may be alleviated.

We believe, using the SBL method to define galaxies or ICL component in the more realistic cosmological hydro-dynamical simulations, e.g., the ILLUSTRIS or EAGLE simulation, the prediction of ICL fraction and the galaxy luminosity function should be more capable to match with observational results. This will be our future work. 


\acknowledgments

The authors thank the anonymous referee for useful suggestions. 
W.P.L. acknowledges support from the National Key Program for Science and Technology Research and Development (2017YFB0203300). 
W.P.L. and X.K. acknowledge supports by the NSFC projects (No. 11473053, 11233005, U1331201, 11333008) 
and National Key Basic Research Program of China (No. 2015CB857001, 2013CB834900). 
X.K. is also supported by the foundation for Distinguished Young Scholars of Jiangsu Province (No. BK20140050). 
The simulations were run in the Shanghai Supercomputer Center and the data analysis was performed on the supercomputing platform of Shanghai Astronomical Observatory. WC is supported by the {\it Ministerio de Econom\'ia y Competitividad} and  the {\it Fondo Europeo de Desarrollo Regional} (MINECO/FEDER, UE) in Spain through grant AYA2015-63810-P as well as the Consolider-Ingenio 2010 Programme of the {\it Spanish Ministerio de Ciencia e Innovaci\'on} (MICINN) under grant MultiDark CSD2009-00064.

\tabletypesize{\footnotesize} \tabcolsep=0.05cm
\begin{deluxetable*}{lccccccccc} \tablecolumns{10}
	\tablewidth{0pt}
	\tablecaption{The observation and analytic data}
	
	\tablehead{\colhead{Cluster/analytic} & \colhead{$M(10^{14}M_{\odot}$)} & \colhead{$z$} & \colhead{band} & \colhead{method} & \colhead{$\mu_{limit}$}  & \colhead{ $f_{ICL}$}  & \colhead{Reference}
	}
	
	\startdata
	Observational results:
	\\Abell 2744   & 17.6$^{(1)}$ & 0.308 &  F106W & SB profile & Non & 0.08$^{(3)}$   & \cite{Morishita16} 
	\\                    &                     &           &  J           & SB limit    & 25    & 0.051$^{(4)}$ & \cite{MT14}   
	\\MACS 0416 & 21.8$^{(1)}$ & 0.396  & F106W & SB profile & Non & 0.22$^{(3)}$   & \cite{Morishita16} 
	\\MACS 0717 & 65.0$^{(1)}$ & 0.548  & F106W & SB profile & Non & 0.13$^{(3)}$   & \cite{Morishita16} 
	\\MACS 1149 & 57.3$^{(1)}$ & 0.544  & F106W & SB profile & Non & 0.08$^{(3)}$   & \cite{Morishita16} 
	\\C1 0024+17 & 1.5$^{(2)}$ & 0.395  &F625W &SB profile & Non & 0.35$^{(4)}$ & \cite{Jee10}                   
	\\Nearby clusterI & $0.6\sim10.0^{(1)}$ & $0.03\sim0.13$ & SDSS-i & SB profile & Non & $0.3\sim0.5^{(4)}$ &\cite{Gonzalez07}  
	\\A2029 & 11.2$^{(2)}$ & 0.0767  & R & SB profile & Non & 0.34$^{(4)}$ & \cite{Uson1991} 
	\\A1656 & 9.7$^{(2)}$ & 0.0232  & R & SB profile & Non & 0.5$^{(4)}$ & \cite{Bernstein1995} 
	\\A1689 & 9.6$^{(2)}$ & 0.18  & V & SB profile & Non & 0.3$^{(4)}$ & \cite{TF1995} 
	\\A1651 & 6.2$^{(2)}$ & 0.084  & I & SB profile & Non & 0.13$^{(4)}$ & \cite{Gonzalez05} 
	\\C1 0024+1652 & 3.2$^{(2)}$ & 0.39  & V & SB profile & Non & 0.15 $^{(4)}$& \cite{Tyson1998} 
	\\A2670 & 2.7$^{(2)}$ & 0.0745  & R & SB profile & Non & 0.3$^{(4)}$ & \cite{SK1994} 
	\\CLASH & $\sim10.0^{(2)}$ & $0.1\sim0.4$  & B & SB limit & 25 & $\sim0.23$ $^{(4)}$ & \cite{Burke15} 
	\\MACS 1026 & 14.1$^{(2)}$ & 0.44 &  V & SB limit & 26.5 & 0.125$^{(4)}$ & \cite{Presotto14} 
	\\SDSS & $\sim1.0/h^{(2)}$ & $0.2\sim0.3$ &  r & SB limit & 27.5 & $\sim0.109$ $^{(4)}$ & \cite{Zibetti08} 
         \\           & $ \sim5.7^{(2)}$   & $0.15\sim0.4$   &X-ray  & SB profile & Non & $0.2\sim0.4^{(4)}$ & \cite{Budzynski14} 
	\\A2390 &14.9$^{(2)}$ & 0.232  & B & SB limit & 24.75 & 0.16$^{(4)}$ & \cite{Vilchez-Gomez1994} 
	\\A1914 & 9.4$^{(2)}$ & 0.1712  & V & SB limit & 26.5 & 0.15$^{(4)}$ & \cite{Feldmeier04a} 
	\\A1413 & 6.4$^{(2)}$ & 0.1427 &  V & SB profile & Non & 0.13$^{(4)}$ & \cite{Feldmeier02} 
	\\Nearby clusterII & $2\sim40/h^{(2)}$ & $0.05\sim0.3$ &  B & $0.25R_{vir}$ & Non & $0.04\sim0.2^{(4)}$ & \cite{KB07} 
	\\M87 & 0.024$^{(2)}$ &  0.00436  & Non  & ICPNs & Non & 0.12$^{(4)}$ & \cite{Longobardi15} 
	\\M81 Group & 0.012$^{(2)}$ & -0.000113  & Non &  ICRGB & Non & 0.013$^{(4)}$ & \cite{Feldmeier04b} 
	\\Virgo & 1.5$^{(2)}$ & 0.004  & Non & ICPNs & Non & 0.16$^{(4)}$ & \cite{Feldmeier04c} 
	\\ \hline
	Theoretical results:
	\\N-body & $0.01\sim1/h^{(2)}$ & 0  & Non & FoF method & Non & $\sim0.25^{(4)}$ & \cite{Contini14}
	\\GADGET-3 & $0.01\sim10/h^{(1)}$ & 0  & Non & dynamical method & Non & $0.6\sim0.8^{(4)}$ & \cite{Cui2014a}
	\\ &  &  & V & SB limit & 25.5 & $0.3\sim0.4^{(4)}$
	\\ &  &  & V & SB limit & 26.5 & $0.2\sim0.3^{(4)}$
	\\ &  &  & V & SB limit & 27.5 & $0.1\sim0.2^{(4)}$
	\\N-body & $1\sim10/h^{(1)}$ & 0 & Non & FoF method & Non & $0.01\sim0.5^{(4)}$ & \cite{Martel12}
	\\GADGET/-2 & $1\sim10/h^{(2)}$ & 0 & Non & Binding Energy & Non & $0.15\sim0.30^{(4)}$ & \cite{Rudick11}
	\\ & & Non & Non & Instantaneous Density & Non & $0.10\sim0.15^{(4)}$
	\\ & & Non & Non & Density History            & Non & $0.15\sim0.20^{(4)}$
	\\ & & Non & V     & SB limit                        & 26.5 & $0.09\sim0.13^{(4)}$
	\\N-body & $\sim1/h^{(2)}$ & 0  & V & SB limit & 26.5 & $0.1\sim0.15^{(4)}$ & \cite{Rudick06a}
	\\hydr-simulation & $0.01\sim1/h^{(1)}$ & 0 & Non & SB profile & Non & $\sim0.45^{(4)}$ & \cite{Puchwein10}
	\\hydr-simulation & $1/h\sim10^{(2)}$ & 0 & Non & SKID & Non & $0.2\sim0.5^{(4)}$ & \cite{Murante04}
	\\
	\enddata
	\tablecomments{1: $M_{500}$, 2: $M_{200}$, 3: ICL mass fraction, 4: ICL light fraction. The fraction of M87, M81 and Virgo actually is ratio of ICL/(BCG+ICL) luminosity based upon their data.}
	\label{tab:fit}
\end{deluxetable*}



\end{document}